\documentclass[conference,draftclsnofoot,onecolumn]{IEEEtran}%
\usepackage[utf8]{inputenc}
 \usepackage[pdftex]{graphicx}
\usepackage[caption=false,font=footnotesize]{subfig}
\usepackage{multirow}
\usepackage{array}
\usepackage{subscript}
\usepackage{rotating}
\usepackage[nocompress]{cite}
\usepackage{titlesec}
\usepackage{enumitem}
\usepackage{flushend}
\usepackage{url}
\usepackage{amsmath}
\usepackage{fixltx2e}

\DeclareGraphicsExtensions{.pdf}
\newcolumntype{P}[1]{>{\centering\arraybackslash}p{#1}}

\titleformat{\section}{\rmfamily\bfseries\fontsize{12}{14}}{\thesection.}{0.5em}%
		{}\relax
\titlespacing{\section}{0pt}{12pt}{12pt}

\titleformat{\subsection}{\rmfamily\bfseries\fontsize{11}{13}}{\thesubsection.}%
{0.5em}{}\relax
\titlespacing{\subsection}{0pt}{11pt}{11pt}

\titleformat{\subsubsection}[runin]{\rmfamily\bfseries\fontsize{10}{12}}%
{\thesubsubsection.}{0.5em}{}[.]
\titlespacing{\subsubsection}{0pt}{10pt}{0.5em}

\begin{document}

\title{Insider Threat Detection Through Attributed Graph Clustering}
\author{\IEEEauthorblockN{Anagi Gamachchi, Serdar Bozta\c{s}}
\IEEEauthorblockA{School of Science, RMIT University, Melbourne,  Victoria, Australia\\
(anagi.gamachchi, serdar.boztas)@rmit.edu.au}
}

\maketitle

This is a pre-publication version of the paper, which appeared in the Proceedings of the $16^{th}$ IEEE International Conference on Trust, Security and Privacy in Computing and Communications, 2017.

\begin{abstract}
While most organizations continue to invest in traditional network defences, a formidable security challenge has been brewing within their own boundaries. Malicious insiders with privileged access in the guise of a trusted source have carried out many attacks causing far reaching damage to financial stability, national security and brand reputation for both public and private sector organizations. Growing exposure and impact of the whistleblower community and concerns about job security with changing organizational dynamics has further aggravated this situation. The unpredictability of malicious attackers, as well as the complexity of malicious actions, necessitates the careful analysis of network, system and user parameters correlated with insider threat problem. Thus it creates a high dimensional, heterogeneous data analysis problem in isolating suspicious users. This research work proposes an insider threat detection framework, which utilizes the attributed graph clustering techniques and outlier ranking mechanism for enterprise users. Empirical results also confirm the effectiveness of the method by achieving the best area under curve value of 0.7648 for the receiver operating characteristic curve. 
\end{abstract}


\section{Introduction}
Insider threat mitigation or finding the enemy hiding within the boundaries of an enterprise network is one of the most critical and complex cybersecurity threats. By looking at the publicly available threat cases \cite{CappelliTheCERT2012} and published literature \cite{Gheyas2016} \cite{survey1}  on insider threat, it is evident that the insider threat can no longer be treated as a data driven problem. It needs to be considered as a combination of data and human behavior driven problem. This drives the requirement for going beyond technical capabilities to understand the unpredictable behavior of the trusted insider. Though there is no particular demographic profile for malicious insiders, there are common characteristics among the three broad categories of insider threat, namely IT sabotage, theft of intellectual property and IT fraud. These characteristics are based on the type of people involved, the motivation for the attack, the time span and the level of damage caused by the attack. In addition several researchers, e.g., Berk et al \cite{Berk2011} and Massberg et al \cite{Maasberg2015}, highlighted the correlation between Capability (Means), Motivation, and Opportunity (CMO/MMO Model) for triggering a malicious insider activity. In addition to disgruntle workplace behavior, users tend to publicize their inside (in-office) experiences online. This includes their interests which have been influenced by positive and negative organizational dynamics. Therefore the insider threat mitigation framework needs to be expanded so that it can capture users' online presence as well. Several other researchers identified various psychological factors which have direct implications on insider threat problem \cite{5168063} \cite{Colwill2009} \cite{Greitzer2010}. 

Based on the above facts it is evident that insider threat mitigation frameworks need to consider heterogeneous data generated from different information sources. Therefore the analysis of such data needs to be considered as a high-dimensional heterogeneous data problem. Among the various methods proposed in the literature for insider threat detection problem we choose to continue our research on graph based approaches as in our previous work \cite{Gamachchi2015} \cite{Gamachchi2017}. Given the complexity of insider threat problem, a direct application of anomaly detection techniques on plain (i.e., unattributed) graphs would not be sufficient. The ``plain" graph based anomaly detection techniques only consider the topological structure while neglecting the associated vertex/edge attribute values. In order to capture topological structure as well as attribute information, we decided to implement anomaly detection techniques associated with attributed graphs. Given the two main anomaly detection approaches on attributed graphs namely, structure based and community based methods \cite{Akoglu2015} we focus on the later approach which aims to identify community outliers. Community outliers find vertices with significantly deviating attribute values within a community. This type of anomaly detection approach would be beneficial in the special case of insider threat problem as some of the behavioural changes can easily be identified when compared with peer level behavioral patterns.    

\noindent \textbf{\textit{Our Contribution}}:In this work we propose the use of attributed graph anomaly detection techniques for malicious activity detection which has not not been explored in depth in the insider threat domain. In addition, we propose several outlier ranking scores which can be used as an indication of possible risk from each individual in an enterprise network. This scoring scheme is based on both topological structure and high dimensional attribute values generated from heterogeneous data sources. The basic framework proposed in this work is depicted in Figure 1. 

\begin{figure}[!t]
  \begin{center}
				\includegraphics[width=0.8\textwidth]{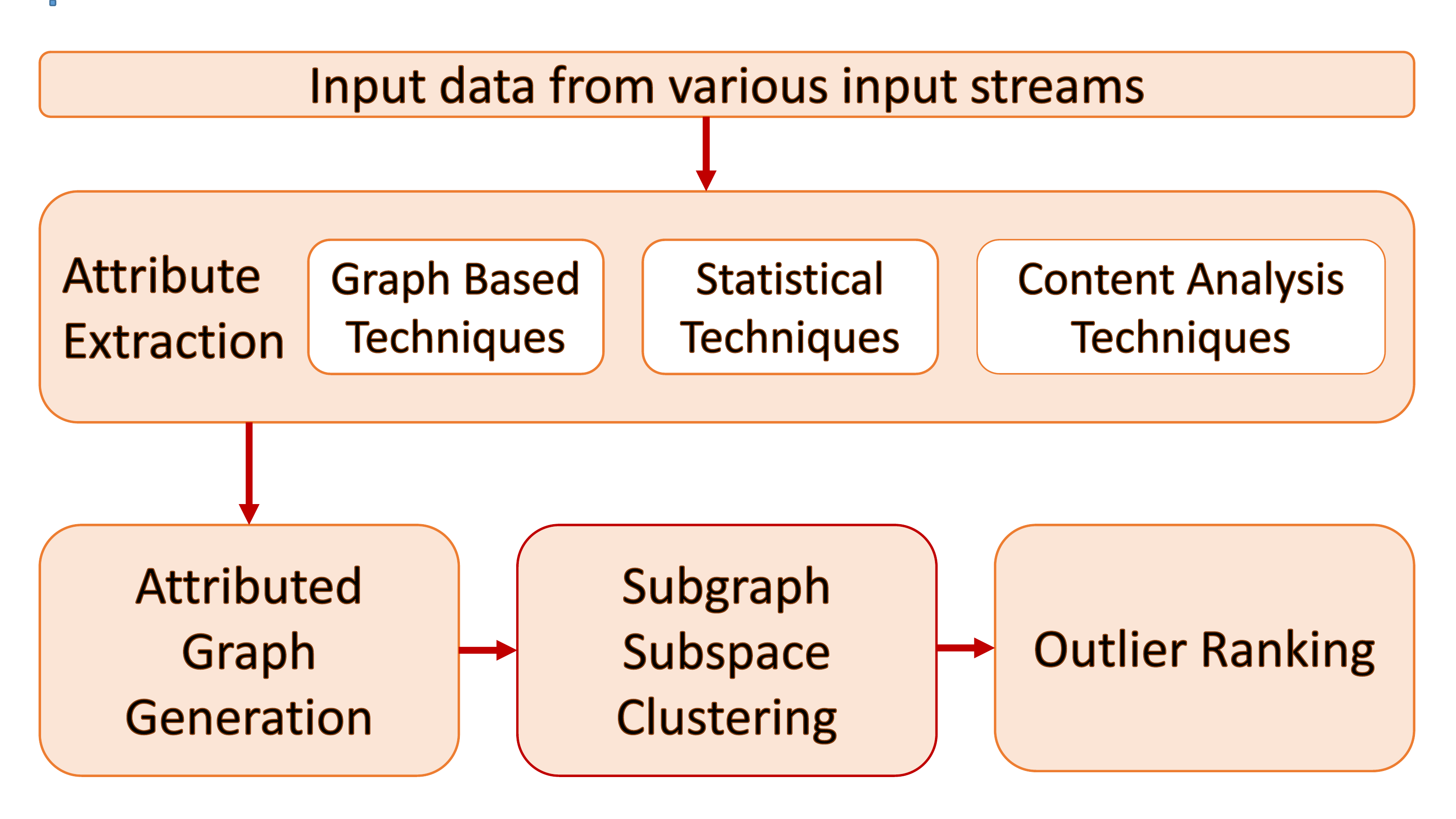}
				\caption{The Proposed Framework for Insider Threat Detection.}
		\label{fig:1}
  \end{center}
	\vspace{-20pt}
\end{figure}

The remainder of this paper is organized as follows. Section $2$ describes the related work on graph based insider threat detection and anomaly detection in attributed graphs. Section $3$ describes the adopted methodology. The experimental results and evaluation of results are discussed in section $4$ and $5$ respectively. Section $6$ concludes the paper indicating conclusions and future directions.

\section{Related Work}
Initial insider threat research efforts were influenced by the techniques used to detect external attacks. Intrusion Detection System (IDS) approaches, system-call-based approaches, data-centric approaches, honeypot approaches, machine learning techniques and visualization methods are the main approaches \cite{survey1} popular among different research groups who are actively engaged in insider threat problem. But many of the above methods led to a high number of false positives and faced several other challenges such as difficulties in handling large datasets, the inability of detecting real-time insider threats, and inaccuracy of the user profiles. 
Emphasizing the significance of bringing together many aspects of insider threat (including technical and behavioural events/indicators as well as as human factors such as precipitating event, personality characteristics, historical behavior, motivation to attack and the capability to attack \cite{Brdiczka2012} \cite{Nurse2014}), insider threat research community focused their attention on high dimensional, heterogeneous data analysis techniques. To overcome the difficulties in analyzing high dimensional, heterogeneous data, some of the insider threat related researchers focused on machine learning approaches and the graph based approaches. Multidimensional data representation capabilities, improved visualization capabilities and feasibility of unsupervised anomaly detection capabilities are the main drivers for us to continue our insider threat detection research on graph based approaches.

Based on the fact that this research work is focused on anomaly detection in attributed graphs on insider threat domain, only the literature related to graph based insider threat detection models, anomaly detection in attributed graphs and outlier ranking in attributed graphs are discussed in the following subsections. 

\subsection{Graph Based Insider Threat Detection Approaches}
The framework proposed by Chen and Malin \cite{chen2011detection} is a graph based approach for insider threat detection in collaborative information systems (CIS). Their model comprises of a relational pattern extraction component and an anomaly detection unit. CIS access logs are mined for communities of users using bipartite graph mapping and dimensionality reduction models. The community based anomaly detection system (CADS) of the proposed framework discovers the nearest neighbors of each user and calculate the deviation of each user to their nearest neighbors. Even though they have identified the requirement of community detection for identifying possible anomalous users, graph mapping is not expanded upto the attributed graphs.

The framework suggested by Eberle and Holder \cite{Eberle2009} is another graph-based approach for malicious insider threat detection. Anomalies are detected using graph substructures which are not isomorphic to graph's normative substructure, by using the minimum description length principle for detecting anomalous activities. They have focused on three broad types of graph anomalies identified in their work namely, insertions, modifications and deletions. Consideration of many attributes from insider threat related inputs is missing in this approach and the usage of multiple algorithms for different anomalies might lead to a complex threat detection framework.

Proactive insider threat detection framework proposed by Brdiczka et al \cite{Brdiczka2012} utilizes graph learning and psychological modeling of users. This model is a combination of structural anomaly detection and psychological profiling model which explore the possibility of including dynamical properties of nodal attributes. Althebyan and Panda \cite{Althebyan2007} have also suggested the use of graph theory to formulate two components, knowledge graph and object dependency graphs. A knowledge graph represents knowledge units for a given insider and they are updated over the time. A dependency graph is a global hierarchical graph that shows all dependencies among various objects. Even though this model tries to include accumulated knowledge of the insider over time on systems and objects it can be improved by including several other parameters such as user's behavioral patterns and psychological aspects.

Another study performed by Nance and Marty \cite{Nance2011} introduced the use of bipartite graphs for identifying and visualizing insider threat. They tried to establish acceptable insider behavior patterns based on workgroup role classifications. High false positive rates is one of the weaknesses of this method even though it is capable of detecting certain insider threats. Kent et al \cite{Kent2015} proposed the use of authentication subgraphs for analyzing users behavior within an enterprise network utilizing a set of subgraph attributes in user profiling. Time series analysis of subgraphs and use of bipartite graphs are also introduced in their work, which is focused on a much more comprehensive analysis in their ongoing work.   

Contributions by above described research work clearly indicate the feasibility of usage of graph based approaches for representation of insider threat related data. Also, it is clear researchers were focused on various anomaly detection techniques for identifying possible malicious users. But many of those proposed methods do not expand to capture the heterogeneity and high dimensionality associated with the insider threat problem. Finally, the above proposed graph based anomaly detection techniques do not consider both graph topology and graph attributes simultaneously.
   
\subsection{Graph Clustering in Attributed Graphs}
Attributed graphs provide a way of obtaining richer graph representation, in which nodes and edges exhibit their properties through both graph topology and graph attributes. These types of graphs can be applicable in many applications such as social networks, transaction networks, technological networks, biological networks, and so on in which most of the related information is stored as attributes of corresponding vertices and edges. Considering the diversity of associated input information with insider threat problem, we believe attributed graphs can be regarded as a appropriate means for data representation. 

Anomaly detection in attributed graphs can be categorised into structure based methods and community based methods \cite{Akoglu2015}. The basic idea behind structure based techniques is to identify uncommon substructures based on the graph connectivity as well as attributes. Community based anomaly detection techniques focus on the nodes in which attribute values are significantly deviate from the other members of the community that they belong to. Based on the reasoning behind above anomaly detection techniques, we believe community based anomaly detection in attributed graphs is more suitable for the insider threat detection frameworks as it would be a better approach for analysing individual behaviour with respect to peers.

Attributed graph clustering has received much more attention in the recent past with the identification of the requirement of analysing high dimensional, heterogeneous data. The method proposed by Zhou et al \cite{Zhou} generates new vertices (``attribute vertices") for each attribute and new edges (``attribute edges") between the vertex and the attribute vertex, if the corresponding vertex has the selected attribute value. Then they use a unified neighborhood random walk model on the attributed-augmented graph to find clusters.

In contrast to the above method Moser et al \cite{COPAM}, G{\"{u}}nnemann et al \cite{Gunnemann2010} \cite{gunnemann2013} proposed attributed graph clustering as a twofold clustering technique which is simultaneously represent attribute subsapces and dense subgraphs. A subspace cluster is a set of objects with an appropriate dimension, in which object attributes are very similar to each other. Dense subgraphs are set of nodes which are densely connected with each other based on ``quasi-clique" property \cite{Liu2008}. Among the above mentioned attributed graph clustering algorithms, we believe subspace and subgraph clustering techniques are more suitable in the insider threat context, in comparison to attribute augmented graph clustering.
 
\subsection{Outlier Ranking Methods}
The algorithms mentioned in the above subsections are mainly focused on graph clustering on attributed graphs. But the scope of these algorithms do not cover the anomaly detection and outlier ranking simultaneously. The methods proposed by Gao et al \cite{Gao} simultaneously finds communities as well as identifying community outliers using an unsupervised learning algorithm called ``CODA". But the usage of global attribute space for community detection would create limitations on direct application on insider threat problem. Another recent clustering mechanism is ``FOCUSCO", proposed by Perozzi et al \cite{Perozzi}, which couples both attributed graph clustering and outlier detection and utilizes user-oriented attribute selection technique which is fairly different from other approaches.

``GOutRank" is the first approach for outlier ranking in subspaces of attributed clustering \cite{Muller2013}. They have used the existing techniques for the selection of subgraphs and subspaces. The proposed outlier ranking mechanism utilize three indicators which includes subspace dimension, cluster dimension and the graph structure. Normalized degree of a node and the normalized eigenvector centrality have been used as graph structure indicators in outlier ranking.

In our opinion, outlier ranking in the context of insider threat problem can be coupled with outlier ranking with subspace/subgraph clustering in attributed graphs. Thus we focus our analysis based on ``GOuRank" outlier ranking mechanism with ``EDCAR" and ``GAMER" algorithms which outstanding in subspace and subgraph clustering in attributed graphs.

\section{Methodology}
This section mainly discusses the adopted methodology in ranking individual users under the proposed insider threat mitigation framework. Due to the lack of availability of proper insider threat datasets we have utilized the insider threat dataset published by CERT at Carnegie Mellon University for this research \cite{CERTDataset}. Out of the different versions of available datasets, the set \textit{``R4.2.tar.bz"} has been used for this analysis. This dataset consists of users' LDAP records and six other broad types of data records (HTTP, logon, device, file, email and psychometric) of 1000 employees of 42 different work roles spanning over a 17 months period. 

As the first step of our approach, the relationships among enterprise users are mapped into an undirected, unweighted graph $G(V,E,A)$ where $V$ is the set of vertices (users), $E$ is the set of edges, and $A$ is the set of attributes. Construction of the graph, based on user relationships is carried out as follows;

\begin{enumerate}
	\item Users are represented as vertices 
	\item Edges between vertices are built based on 
			\begin{itemize}
				\item organizational hierarchy (represent supervisor - subordinate relationship)
				\item email communications (if there is an email communication between two users)
		\end{itemize}
\end{enumerate}

When generating the graph structure, the employees ``supervisor - subordinate" relationship is first mapped in to an undirected edge between corresponding users. In addition, we have used email communication logs to capture users friendship network within the enterprise network. This relationship is captured by analyzing all emails address ``TO", ``CC" or ``BCC" within the enterprise domain. Then, an edge between the sender and the recipient is created. Within the scope of this research, we exclude the directionality and the weights of edges as subspace and subgraph clustering algorithms have not been developed to apply on directed graphs. Also we have excluded relationships with external users in the generated graph, but we have captured the relationship with external users as a separate attribute. 

In addition each vertex $v$ (in this work this is a ``user") is described by a vector $(a_{1},\ldots,a_{d}) \in \Re_{d}$ in a $d- dimensional$ continuous data space where the attributes are denoted $A = (A_{1}, \ldots, A_{d})$. 

\subsection{Attribute Extraction}
In this work statistical techniques are used for extracting related attributes from various information streams such as logon/logoff records, web access records, email communications, removable media usage records and file copy activities. This mainly includes extraction of maximum, minimum and average value of the selected parameters. The list of parameters used in this analysis as well as number of calculated sub-parameters are summarized in Table 1. All categorical attributes of users (role, functional unit, department and team) are mapped into integers and time values are converted to decimal format before feeding into the attributed graph clustering algorithms. Parameter values which correspond to file type are calculated as the ratio between the number of files copied by a user in the selected file type and the total number of files copied by the same user.  

\renewcommand{\arraystretch}{1.1} 
\begin{table}
\centering
\caption{Parameter List }
\begin{tabular}{|P{0.1\textwidth}|p{0.6\textwidth}|P{0.1\textwidth}|}
\hline \textbf{Index} & \textbf{Attribute Name} & \textbf{No.of sub attributes}\\
\hline 1 & Number of recipients in TO/CC/BCC fields & 9\\
\hline 2 & Email size & 3\\
\hline 3 & Number of attachments & 3\\
\hline 4 & Daily number of sent emails (all/Business Hours (BH)/After Hours (AH) & 9\\
\hline 5 & Email sent time & 3\\
\hline 6 & Number of devices used for email activities & 1\\
\hline 7 & Number of email address & 1\\
\hline 8 & Number of internal users connected through emails & 1\\
\hline 9 & Number of external users connected through emails & 1\\
\hline 10 & Role & 1\\
\hline 11 & Functional unit & 1\\
\hline 12 & Department & 1\\
\hline 13 & Team & 1\\
\hline 14 & Logon Time (all/BH/AH) & 9\\ 
\hline 15 & Logoff Time (all/BH/AH) & 9\\ 
\hline 16 & Daily number of logons (all/BH/AH) & 9\\
\hline 17 & Daily number of logoffs (all/BH/AH) & 9\\
\hline 18 & Daily number of devices used for logon/logoff activities & 3\\
\hline 19 & Number of daily usage - removable media (all/BH/AH) & 9\\
\hline 20 & Removable media usage time (BH/AH) & 9\\ 
\hline 21 & Number of devices used for removable media & 4\\
\hline 22 & Total number of days - removable media & 1\\
\hline 23 & File copy time (all/BH/AH) & 9\\
\hline 24 & Number of days - file copy (all/BH/AH) & 3\\
\hline 25 & Number of files - file copy (all/BH/AH) & 9\\
\hline 26 & File type & 6\\
\hline 27 & Number of devices used for file copy\ & 1\\
\hline 
\end{tabular}
\vspace{-10pt}
\end{table}

\subsection{Subgraph and Subspace Extraction}
In this work we propose the use of community based anomaly detection techniques in attributed graphs. Our aim is to identify the outlier nodes with attribute values which deviate significantly from the rest of the nodes of the selected community \cite{Akoglu2015}. In high dimensional attributed graphs consideration of all set of attributes would not be able to clearly identify the exceptional outliers. Also outliers would not be visible if we consider the irrelevant attributes \cite{Muller2013}. This argument is much more applicable in the insider threat problem as, finding the hidden truth really depends on the analysis of enormous amount of heterogeneous data. Also detection of several insider threat cases need to be analyzed as threat scenarios which involves only a few of the selected attributes or a subset of the global attribute space. 

As proposed in \cite{Muller2013} the first step of outlier ranking in high dimensional, heterogeneous data requires detection of an graph context of an outlier (subgraphs) and relevant attribute set (subspace) in which an outlier is deviating. As mentioned in Section 2 we decided to adopt the subspace clustering algorithms ``EDCAR" \cite{gunnemann2013} and ``GAMER" \cite{Gunnemann2010} for subspace clustering. Due to the fact that subspace and subgraph clustering is an emerging field in graph clustering techniques, we have decided to used the above two algorithms which outperform on other clustering techniques on attributed graphs.

Subspace clustering result of an attributed graph is a set of subspace clusters $R_{es}$ $=$ $\{(C_{1},S_{1}) \ldots (C_{n},S_{n})\}$, where $C_{i} \subset V$ is a densely connected subgraph with high attribute similarity in the subspace $S_{i} \subset A$ \cite{Muller2013}. 

\subsection{Outlier Ranking}
The next step involves ranking of all vertices of the graph $G$ based on the identified subgraphs and subspaces. In the context of insider threat this scoring function can be considered as a user profile scoring system which can be utilized by a security analyst for further investigations. The first outlier ranking approach (GOutRank) in subspaces of attributed graphs was proposed by \cite{Muller2013}. We first calculate the outlier ranking scores based on two proposed scoring mechanisms based on node degree scoring and eigenvalue scoring as mentioned in equations (1) and (2). 

\begin{eqnarray}
     score_{1}(v)\!\!\!\!\!&=&\!\!\!\!\!\! \frac{1}{3} \!\! \cdot \!\!\! \sum_{\{(C,S)\in R_{es} \lvert v \in C\}} \! \frac{\lvert C\rvert}{c_{max}} \! + \! \frac{\vert S\rvert}{s_{max}} \!+\! \frac{\vert deg(v)\rvert}{deg_{max}} 
		\\
     \label{eq:3}
   score_{2}(v)\!\!\!\!\!\!&=&\!\!\!\!\!\! \frac{1}{3} \!\! \cdot \!\!\! \sum_{\{(C,S)\in R_{es}  \lvert v \in C\}} \! \frac{\lvert C\rvert}{c_{max}} \! + \! \frac{\vert S\rvert}{s_{max}} \!+\!  \frac{\vert EC(v)\rvert}{EC_{max}}
     \label{eq:4}
\end{eqnarray}

In addition to above two scoring mechanisms we propose the use of betweenness centrality measure which is a measure of centrality in graphs based on shortest paths. The corresponding score is calculated as in equation $(3)$.  

\begin{eqnarray}
     \!\!score_{3}(v)\!\!\!\!\!\!&=&\!\!\!\!\!\! \frac{1}{3} \!\! \cdot \!\!\! \sum_{\{(C,S)\in R_{es} \lvert v \in C\}} \frac{\lvert C\rvert}{c_{max}} \! + \! \frac{\vert S\rvert}{s_{max}} \!+\! \frac{\vert BC(v)\rvert}{BC_{max}}
\label{eq:5}
\end{eqnarray}

Also we calculate $score_{4}(v)$, $score_{5}(v)$ and $score_{6}(v)$, which are combinations of above mentioned 3 graph properties (degree, eigenvector centrality and betweeneess centrality) as in equations (4), (5) and (6).

\begin{eqnarray}
  	score_{4}(v)\!\!\!\!\!\!&=&\!\!\!\!\!\! \frac{1}{4} \!\! \cdot \!\!\! \sum_{\{(C,S)\in R_{es} \lvert v \in C\}} A  +  \frac{\vert EC(v)\rvert}{EC_{max}}\\
    \label{eq:6}
		score_{5}(v)\!\!\!\!\!&=&\!\!\!\!\!\! \frac{1}{4}  \!\! \cdot \!\!\! \sum_{\{(C,S)\in R_{es} \lvert v \in C\}} A + \frac{\vert BC(v)\rvert}{BC_{max}}\\
    \label{eq:7}
		score_{6}(v)\!\!\!\!\!&=&\!\!\!\!\!\! \frac{1}{5}  \!\! \cdot \!\!\!\sum_{\{(C,S)\in R_{es} \lvert v \in C\}} A + \!\!\frac{\lvert EC(v)\rvert}{EC_{max}} \!\! + \!\! \frac{\vert BC(v)\rvert}{BC_{max}}
    \label{eq:8}
\end{eqnarray}

where
\[
A=\frac{\lvert C\rvert}{c_{max}} + \frac{\lvert S\rvert}{s_{max}} + \frac{\lvert deg(v)\rvert}{deg_{max}}
\]

In all the above equations ${\lvert C \rvert}$ is the number of objects in cluster $C$ and ${\lvert S \rvert}$ is the number of attributes in the subspace. $C_{max}$ and $S_{max}$ are the maximal cluster size and the maximal dimensionality in $R_{es}$. Also ${deg(v)}/{deg_{max}}$, ${EC(v)}/{EC_{max}}$ and ${BC(v)}/{BC_{max}}$ represent normalized edge degree, eigenvector centrality and betwenness centrality respectively.

Vertices not clustered or clustered as part of very small, sparsely connected communities with low dimensional subspaces are considered as clear outliers. Therefore the vertices with lower score values can be considered as outliers while vertices with higher score values can be treated as normal.

\section{Experimental Results}
The figure 2(a) corresponds to the user relationship graph which is generated using both organizational hierarchy and email communications within the enterprise network. For comparison purposes, we have illustrated the user relationship graph generated considering only the organizational structure. Less dense graph {2(b)} indicates that we can not extract much structural information about user relationships just considering the organizational structure. Therefore we believe that the graph generated with both organizational hierarchy and email communications would be a good choice for identifying inter user relationships. The graph 2(a) consists of $1000$ vertices and $116,097$ undirected and unweighted edges. Each vertex is defined as a vector of $125$ attributes, which were extracted from different information sources summarized in Table 1.

\begin{figure}[!t]
\centering
\subfloat[Organizational Hierarchy and Email Communication]{\fbox{\includegraphics[scale=0.4]{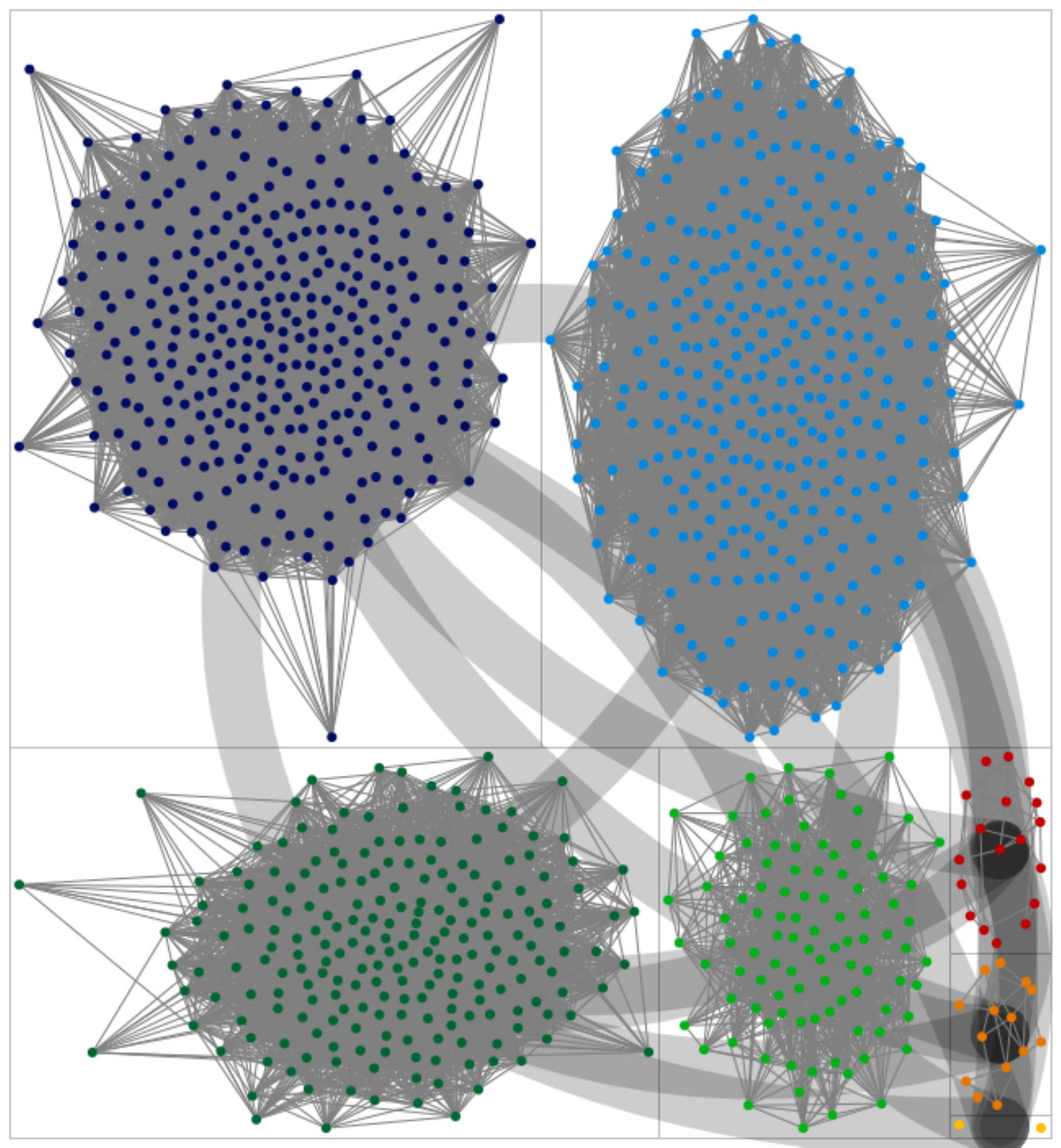}}%

\label{fig_2_1}}
\hfil
\subfloat[Organizational Hierarchy]{\fbox{\includegraphics[scale=0.4]{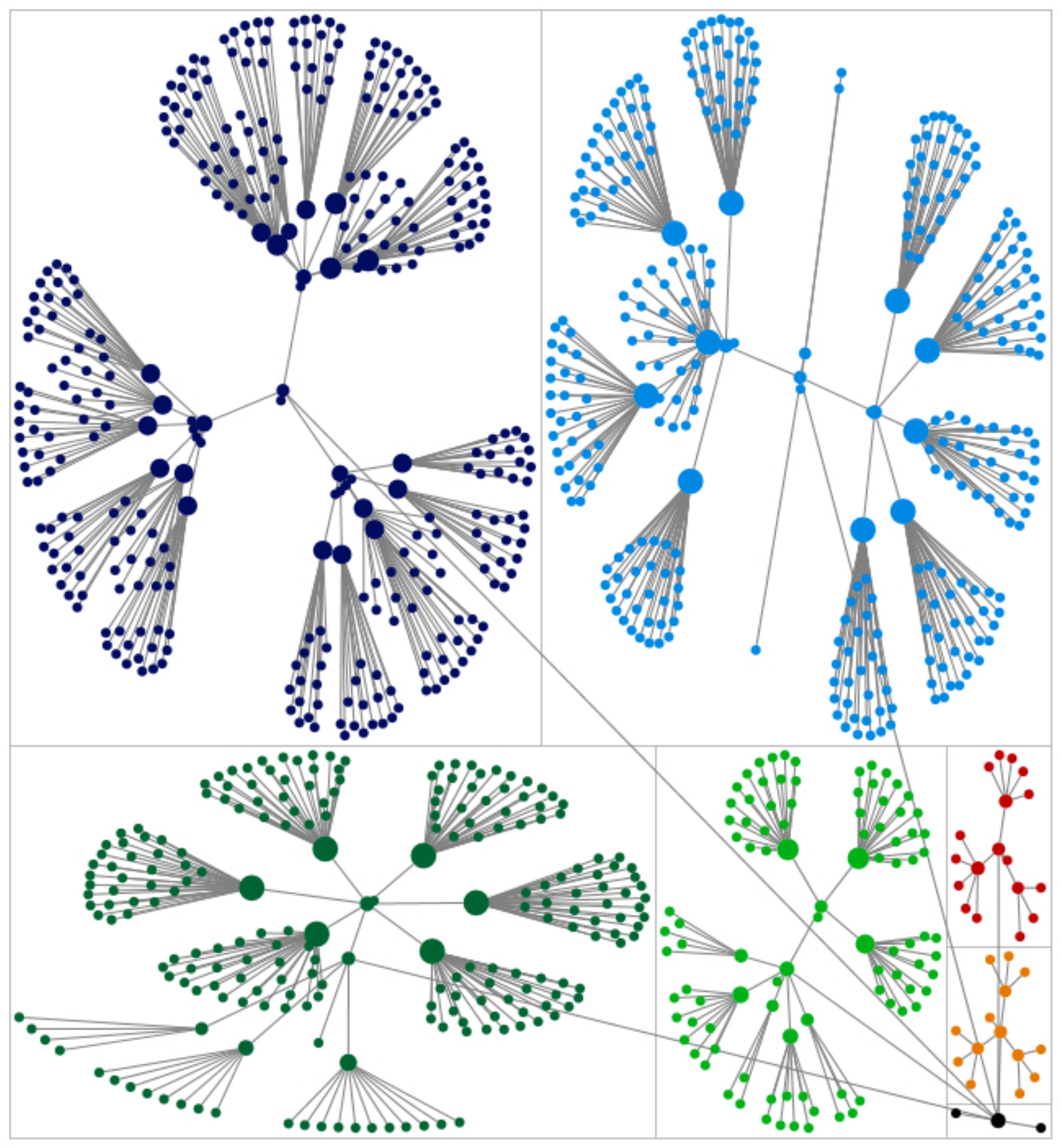}}%
\label{fig_2_2}}
\hfill
\caption{User Relationship Graphs}
\label{fig_sim}
\end{figure}

\begin{table}
\centering
\caption{Clustering Summary of EDCAR Algorithm}
\begin{tabular}{|P{0.1\textwidth}|P{0.1\textwidth}|P{0.1\textwidth}|P{0.1\textwidth}|P{0.1\textwidth}|P{0.1\textwidth}|P{0.1\textwidth}|}
\hline \multirow{3}{*} {\textbf{n\textsubscript{min}}} & \multicolumn{3}{c|}{\textbf{No.of Clusters}} & \multicolumn{3}{c|}{\textbf{No.of Clustered Users}}\\
\cline{2-7}  & \multicolumn{3}{c|}{\textbf{s\textsubscript{min}}} & \multicolumn{3}{c|}{\textbf{s\textsubscript{min} }}\\
\cline{2-7}  & \textbf{3} & \textbf{4} & \textbf{5} & \textbf{3} & \textbf{4} & \textbf{5}\\
\hline 2 & 237 & 118 & 66 & 745 & 493 & 342\\
\hline 3 & 248 & 114 & 83 & 763 & 474 & 429\\
\hline 4 & 240 & 121 & 55 & 734 & 501 & 282\\
\hline 5 & 234 & 111 & 81 & 719 & 480 & 417\\
\hline 6 & 250 & 120 & 67 & 768 & 505 & 339\\
\hline 7 & 251 & 109 & 84 & 766 & 461 & 426\\
\hline 8 & 248 & 113 & 90 & 761 & 476 & 461\\
\hline 9 & 241 & 104 & 89 & 732 & 434 & 456\\
\hline 10 & 238 & 115 & 66 & 744 & 459 & 336\\
\hline 
\end{tabular}
\end{table}

\begin{table}
\centering
\caption{Clustering Summary of Topological Graph Clustering Algorithms}
\begin{tabular}{|p{0.25\textwidth}|P{0.15\textwidth}|p{0.4\textwidth}|}
\hline \textbf{Clustering Algorithm} & \textbf{No. of Clusters} & \textbf{Cluster Membership Size}\\
\hline Edge Betweeness  & 226 & $ 1 \times 225$, 775\\
\hline Walktrap & 12 & $ 31 \times 11$, 659\\
\hline Multilevel & 11 & 20,23,37,39,76,107,108,135,150\\
\hline Fastgreedy & 3 & 260,298,442\\
\hline Leading Eigenvector & 3 & 245,273,482\\
\hline Informap & 1 & 1000\\
\hline Label propagation & 1 & 1000\\
\hline
\end{tabular}
\end{table}

\begin{figure}[!t]
\centering
\subfloat[Case O (n=3,s=8)]{\fbox{\includegraphics[width=0.45\textwidth]{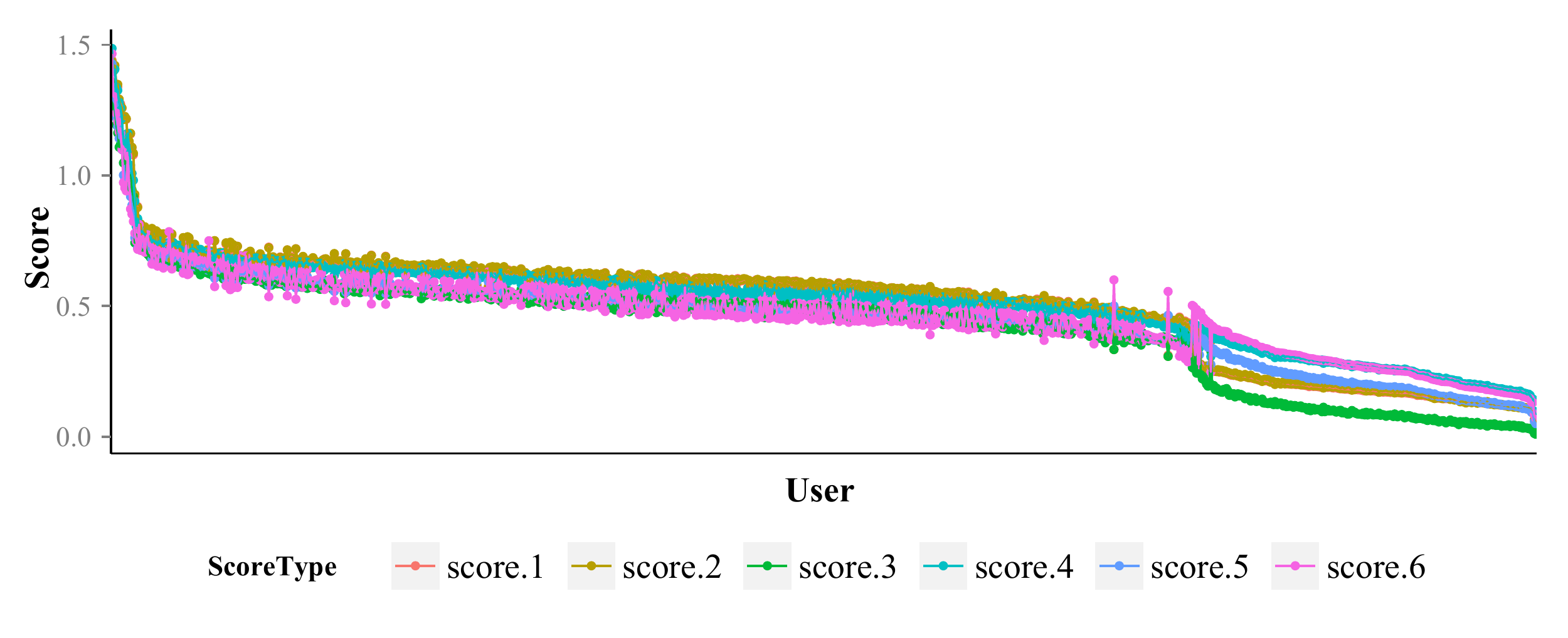}}%
\label{fig_3_1}}
\hfil
\subfloat[Case E (n=4,s=3)]{\fbox{\includegraphics[width=0.45\textwidth]{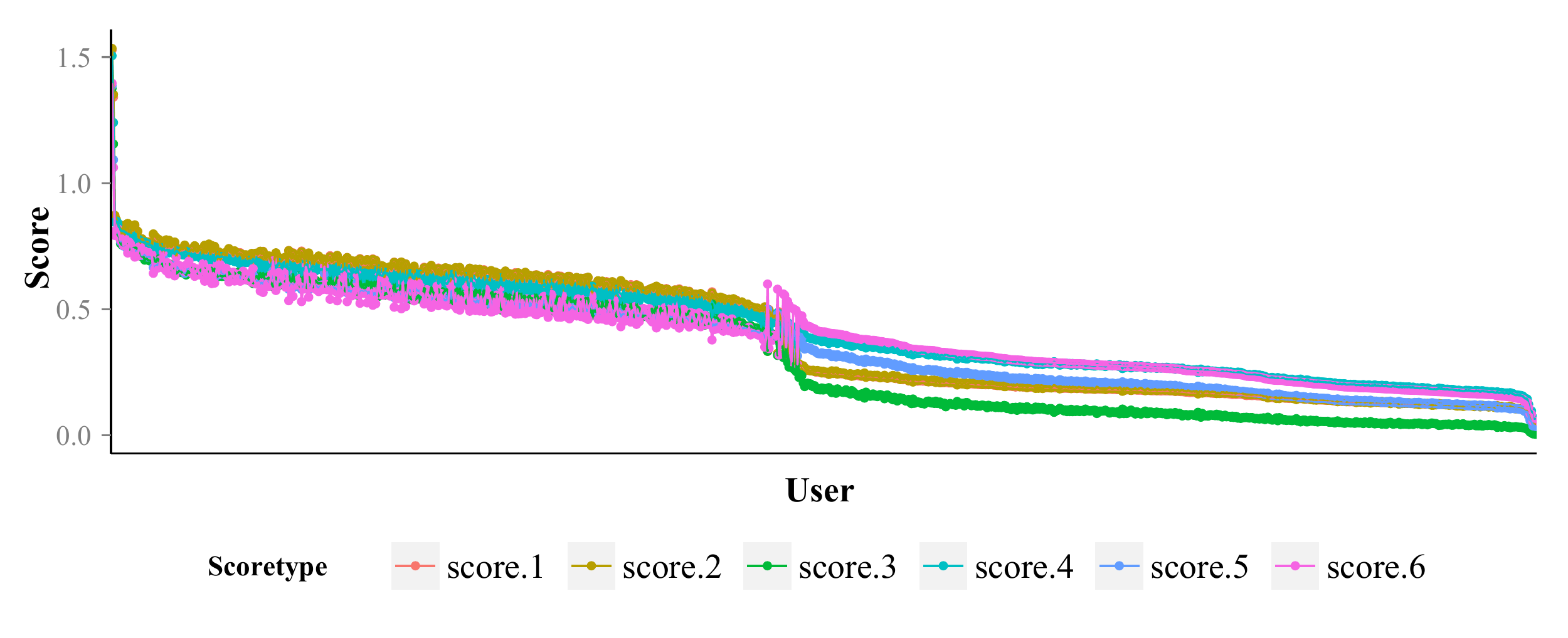}}%
\label{fig_3_2}}
\hfil
\subfloat[Case F (n=5,s=3)]{\fbox{\includegraphics[width=0.45\textwidth]{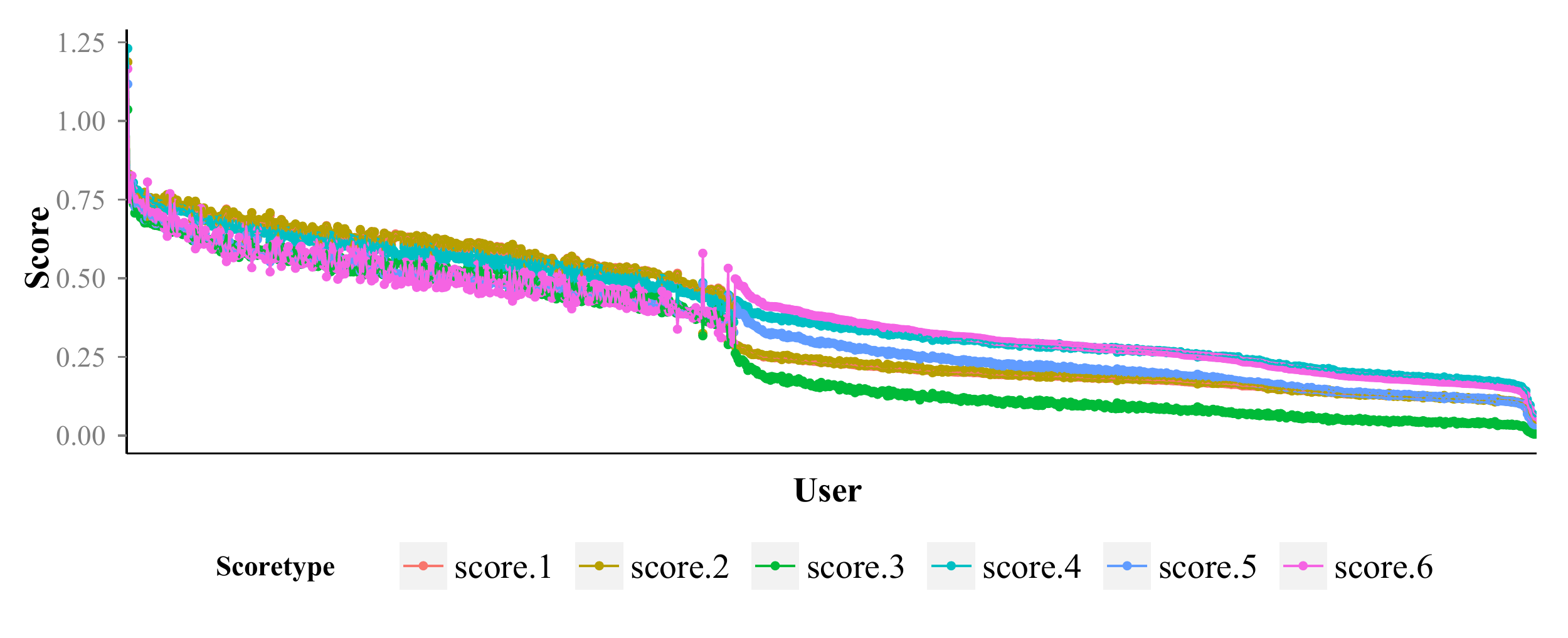}}%
\label{fig_3_3}}
\hfil
\caption{Variation of Outlier Scores for Different Scoring Functions}
\label{fig_OS}
\vspace{-10pt}
\end{figure}

The experiments were carried out with ``EDCAR" \cite{gunnemann2013} and ``GAMER"\cite{Gunnemann2010} algorithms for different values of $n_{min}$ (minimum number of members in a cluster) and $s_{min}$ (minimum subspace dimension) keeping other parameters fixed ($\gamma_{min}=0.5$, $a=b=c=1$, $r_{obj}=0.1$ and $r_{dim}=0.1$). We could not get any results for the tested parameter values with ``GAMER" algorithm within 3 days of processing. Therefore the table 2 summarizes the number of clusters and number of users clustered together using ``EDCAR" algorithms with different $n_{min}$ and $s_{min}$ values. Based on the number of clusters and the total number of users grouped together in each case, we achieved better clustering for lower values of $n_{min}$ for constant $s_{min}$ values. Based on this result we can see $3$ is a good candidate for $n_{min}$ in this context. The total number of users belonging to clusters as well as number of clusters reduces with the increase of the minimum number of members in a cluster ($n_{min})$. This resulted in a higher number of users not belonging to any of the clusters. If the number of users who can be clustered with at least a few of the other employees is higher, this is an indication that they have at least several behavioral similarities. Users who can not be clustered based on both structural and attribute properties can be regarded as high risk employees, as their behavior is somewhat different from others.  

To compare the effectiveness of the attributed graph clustering, we have performed graph clustering on the same graph with several other ``plain" graph clustering techniques. (All these algorithms are available in R in the ``igraph" package). These graph clustering methods mainly looked at the graph structure and did not consider the associated vertex attributes. The number of clusters obtained in selected topological graph community detection techniques is summarized in Table 3. As in the resulted number of clusters in the structural community detection methods we can clearly see many of these algorithms do not suit our problem due to the relatively smaller number of clusters resulting in a large number of unclustered users or clusters with a large number of memberships. Though the number of clusters obtained from the edge betweenness community detection algorithm \cite{Girvan11062002} is relatively close to the number of clusters obtained with attributed graph clustering, many of the clusters (225) comprise of only a single vertex, which can be regarded as not clustered at all and all the other users were clustered in a single community. Therefore it is evident that the application of attributed graph clustering is much more meaningful than the traditional structural community detection techniques in the context of insider threat problem. Also from the two different attributed graph clustering methods ``EDCAR" and ``GAMER", we observe better results for ``EDCAR" as ``GAMER" was not able to produce results within at least 3 days of run time.   

\subsection{Outlier Ranking}	
Outlier ranking scores corresponding to different scoring functions (defined in equations (1) to (6)) are calculated and illustrated in figure 3, in the descending order of scores. (best 3 graphs corresponding to three different $n_{min}$ values are illustrated, as score distributions corresponding to other cases with same $n_{min}$ values also have similar shape of graphs with different values of scores). The distribution of outlier scores in $CASE\_O$, indicates that the majority of users have fairly common ranking scores while a minority of users with either relatively higher scores or lower scores. But based on the subspace and subgraph clustering algorithm, the users with higher scores can be treated as normal as they correspond to densely connected subgraphs for a set of attribute. Therefore the users with the lower scores can be regarded as possibly suspicious actors, who are deviating from the majority of users. The rest of the outlier score graphs indicate that the percentage of users with lower scores increases with the increase of $n_{min}$. Therefore we believe that we achieve better performance by keeping $n_{min}$ or the minimum number of members in a cluster to relatively a small number. This argument can be easily correlated with the real world environments as a few employees will behave similarly when we consider set of associated parameters. 

\begin{figure*}[!t]
\centering
\subfloat[Case O (n=3,s=8)]{\fbox{\includegraphics[width=0.27\textwidth]{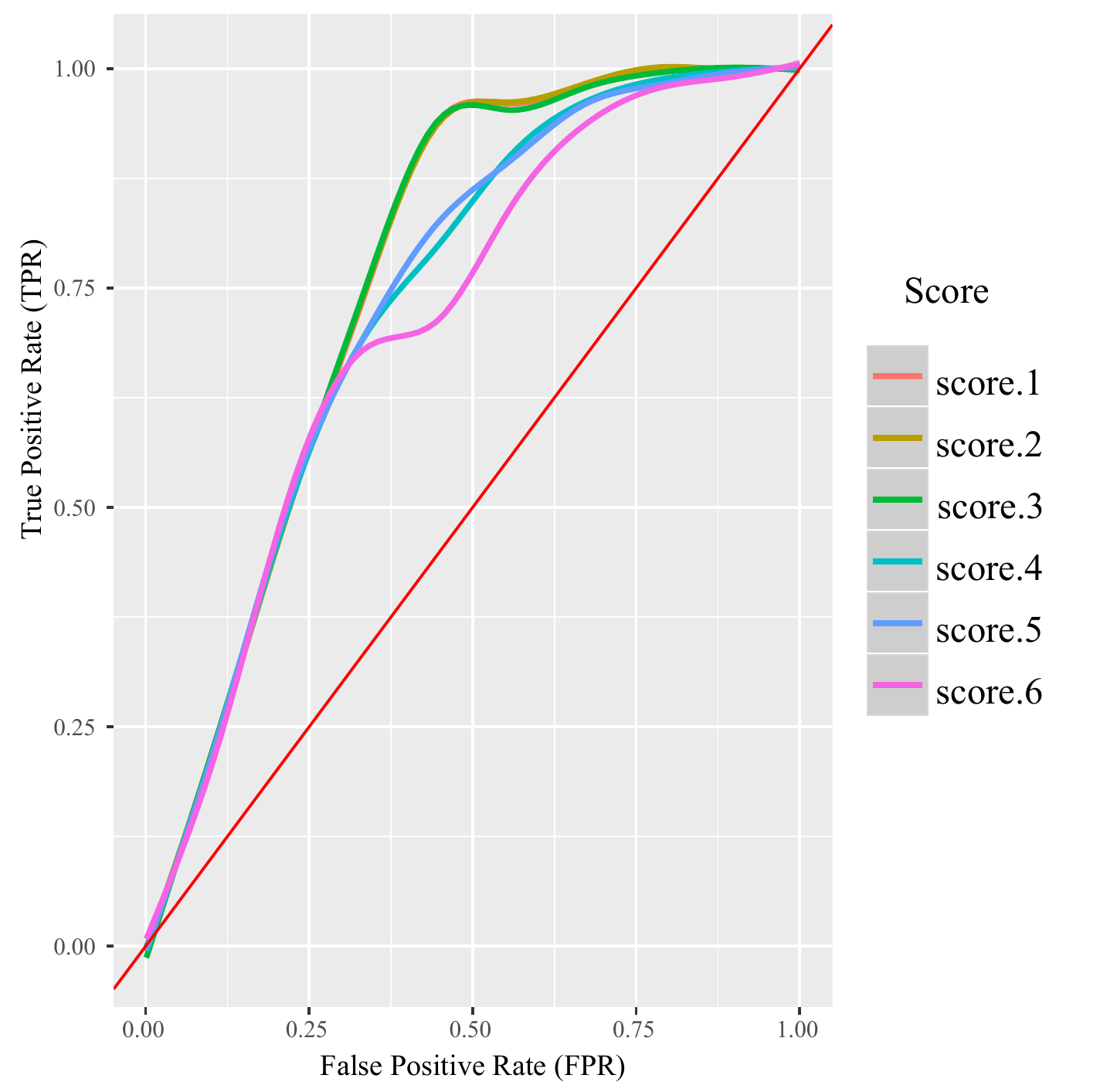}}%
\label{fig_5_1}}
\hfil
\subfloat[Case A (n=3,s=2)]{\fbox{\includegraphics[width=0.27\textwidth]{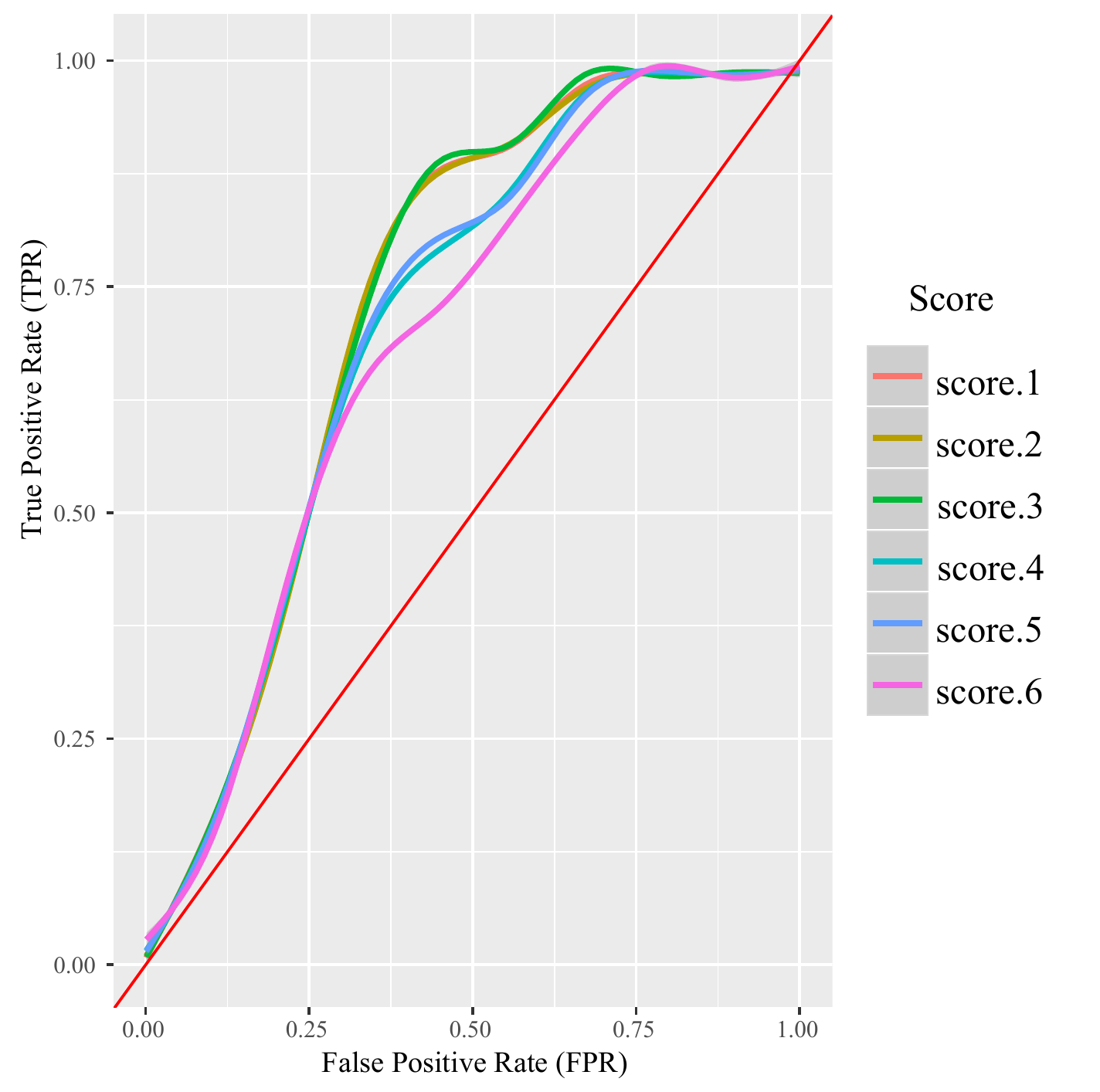}}%
\label{fig_5_1}}
\hfil
\subfloat[Case P (n=3,s=9)]{\fbox{\includegraphics[width=0.27\textwidth]{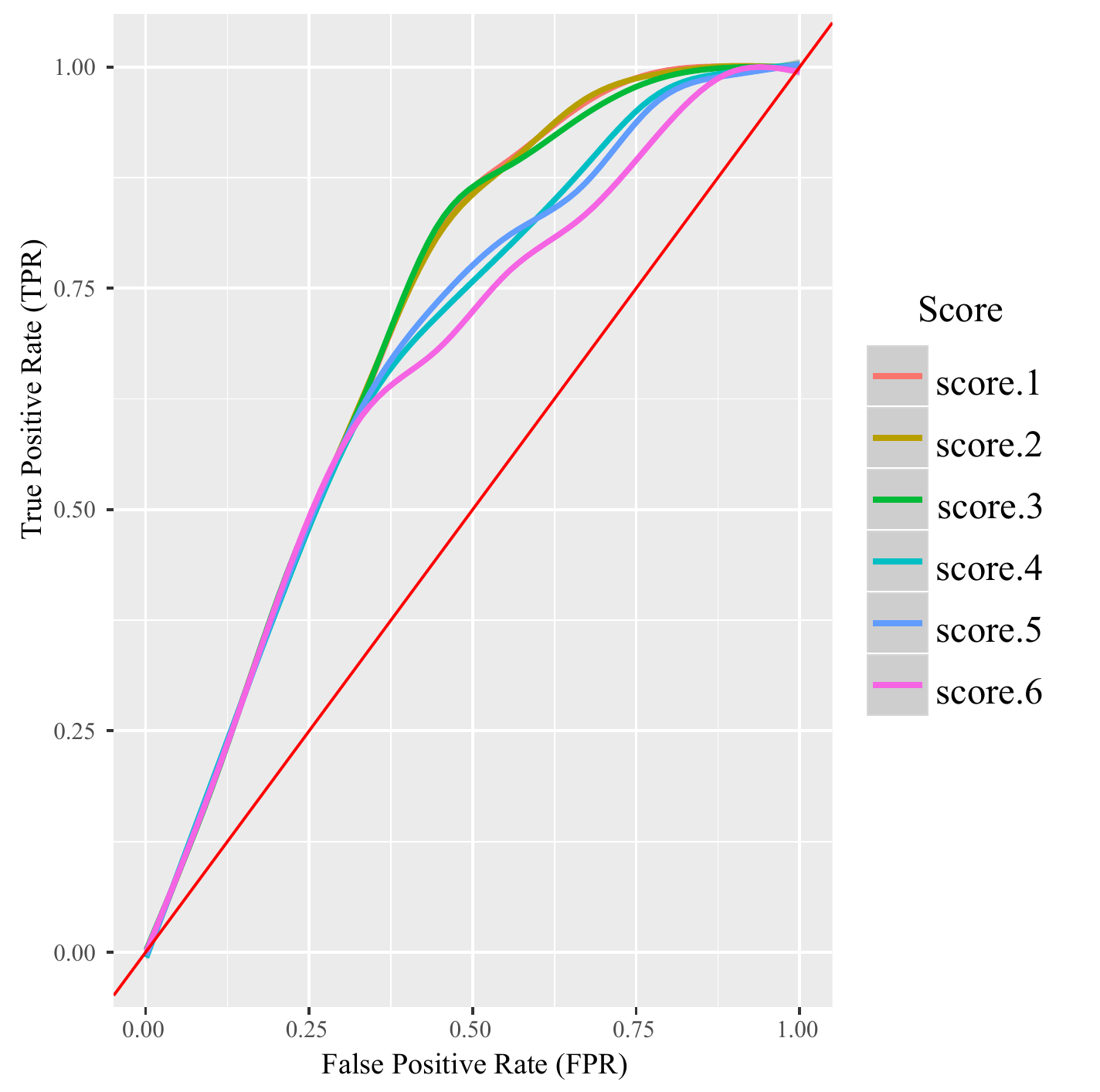}}%
\label{fig_5_1}}
\hfil
\caption{Receiver Operating Characteristic (ROC) curves; the above graphs illustrate the ROC curves for different scoring functions with EDCAR clustering algorithm. The x-axes denote false positive rate (FPR) and the y-axes denote true positive rate (TPR).}
\label{fig_ROC}
\vspace{-5pt}
\end{figure*}

\section{Evaluation of Results}
One of the major challenges faced by the insider threat research community is difficulties and limitations of cross-validation of proposed methods. This is mainly due to the use of different datasets (in most cases private datasets), the focus on a specific problem, or application to a particular scenario when presenting individuals' insider threat models. Thus we have used the ground truth from the dataset for evaluation of the results. The area under curve (AUC) value of the Receiver Operating Characteristic (ROC) curves is used as the indicator for the assessment. Figure 4 illustrates the corresponding ROC curves for each test scenario. Corresponding AUC values for different scoring functions based on ``EDCAR" attributed graph clustering technique are summarized in Table 4 (all test cases for $s_{min}<=5$ and only the test cases for $s_{min}>5$ with $n_{min}=3$ are shown). From the different $n_{min}$ values the best performance is observed for $n_{min}=3$. With the value of $n_{min}=3$ the best performance is achieved with $s_{min}$ of $8$. We achieved the best AUC value $(0.7648)$ of ROC curve for the scoring function defined using node degree measure as in equation (1) with the minimum cluster size of $3$ and the minimum attribute dimension of $8$. The scoring functions defined with individual graph properties (equations (1) to (3)) performed better than the scoring functions defined as a combination of graph properties (equations (4) to (6)). But with this empirical results we could not highlight any of the scoring functions defined in (equations (1) to (3)) as the best performer. But based on the empirical results we believe the proposed framework works fairly well for the insider threat problem. Also we shall continue our research on identifying several other graph structural features that can be used in final scoring functions.

\renewcommand{\arraystretch}{1.1} 
\begin{table*}[!t]
\centering
\caption{The Area Under Curve (AUC) Values for Different Scoring Functions with EDCAR Clustering Algorithm}
\begin{tabular}{|P{0.05\textwidth}|P{0.05\textwidth}|P{0.05\textwidth}|P{0.1\textwidth}|P{0.1\textwidth}|P{0.1\textwidth}|P{0.1\textwidth}|P{0.1\textwidth}|P{0.1\textwidth}|}
\hline \textbf{Case} & \textbf{\textbf{n\textsubscript{min}}} & \textbf{\textbf{s\textsubscript{min}}} & \textbf{score.1} & \textbf{score.2} & \textbf{score.3} & \textbf{score.4} & \textbf{score.5} & \textbf{score.6}\\
\hline A & 3 & 2 & 0.7313 & 0.7287 & 0.7478 & 0.7017 & 0.7090 & 0.6927 \\
\hline B & 4 & 2 & 0.6687 & 0.6690 & 0.6669 & 0.6576 & 0.6572 & 0.6503 \\
\hline C & 5 & 2 & 0.6017 & 0.6014 & 0.6009 & 0.5976 & 0.5974 & 0.5952 \\
\hline D & 3 & 3 & 0.6931 & 0.6923 & 0.6958 & 0.6634 & 0.6656 & 0.6430 \\
\hline E & 4 & 3 & 0.6794 & 0.6800 & 0.6781 & 0.6707 & 0.6704 & 0.6641 \\
\hline F & 5 & 3 & 0.6040 & 0.6045 & 0.6032 & 0.5963 & 0.5968 & 0.5929 \\
\hline G & 3 & 4 & 0.7114 & 0.7109 & 0.7100 & 0.6795 & 0.6801 & 0.6623 \\
\hline H & 4 & 4 & 0.6453 & 0.6453 & 0.6442 & 0.6335 & 0.6341 & 0.6269 \\
\hline I & 5 & 4 & 0.5732 & 0.5742 & 0.5708 & 0.5705 & 0.5695 & 0.5698 \\
\hline J & 3 & 5 & 0.6996 & 0.6994 & 0.7004 & 0.6717 & 0.6728 & 0.6516 \\
\hline K & 4 & 5 & 0.5917 & 0.5923 & 0.5918 & 0.5825 & 0.5826 & 0.5816 \\
\hline L & 5 & 5 & 0.5479 & 0.5474 & 0.5473 & 0.5374 & 0.5391 & 0.5337 \\
\hline M & 3 & 6 & 0.6988 & 0.6967 & 0.7021 & 0.6618 & 0.6669 & 0.6371 \\
\hline N & 3 & 7 & 0.7035 & 0.7025 & 0.7034 & 0.6786 & 0.6789 & 0.6648 \\
\hline O & 3 & 8 & 0.7648 & 0.7649 & 0.7638 & 0.7329 & 0.7349 & 0.7114 \\
\hline P & 3 & 9 & 0.7160 & 0.7153 & 0.7139 & 0.6808 & 0.6814 & 0.6621 \\
\hline Q & 3 & 10 & 0.6743 & 0.6731  & 0.6775  &0.6443  &0.6476  &0.6220  \\
\hline 
\end{tabular}
\vspace{-5pt}
\end{table*}	
	
\section{Conclusions and Future Work}
In this paper, we have introduced the use of attributed graphs for representing high dimensional, heterogeneous data in the context of insider threat detection. The insider threat detection framework proposed here combines attributed graph clustering techniques and outlier ranking in subspaces of attributed graphs. To the best of our knowledge, though the graph based approaches are adopted in malicious insider threat detection frameworks, anomaly detection in attributed graphs using subspace and subgraph clustering has not been widely applied in insider threat detection frameworks. Thus this is one of the early attempts at using subspace/subgraph clustering coupled with outlier ranking for anomaly detection in the insider threat research domain. We have utilized two main subspace/subgraph clustering algorithms namely ``EDCAR" and ``GAMER" for community detection in attributed graphs. Also, we have adopted the outlier ranking mechanism ``GOutRank", which is the first approach for outlier ranking in subspaces of attributed graphs. In addition to the proposed vertex scoring mechanisms in ``GOutRank" algorithm, we also identified betweenness centrality measure as another useful centrality score. Based on the experimental results we obtained the best AUC value of ROC curves for ``EDCAR" clustering algorithm with outlier ranking score defined based on the node degree.  

In subsequent work, we will look at more graph based properties such as other centrality measures as well as subgraph properties to incorporate in outlier ranking scoring mechanism. Also we focus our attention on dynamic attributed graph clustering, which can be useful in integrating the time dimensionality of the problem. Integration of scenario based attribute subspaces would also be another path we would like to focus in the future. Also our continuous work focuses on a comprehensive parameter list related to insider threat problem and related attribute extraction techniques in addition to our previous work \cite{Gamachchi2015} \cite{Gamachchi2017}. This would include content analysis techniques which will be useful in identifying web access behavior as well as email communications. On the other side we would also look at attribute extraction methodologies from social and professional networks, which can be used as an evidence for the hidden behavior of a user.

\section*{Acknowledgments}
The authors were supported by an Australian Research Council (ARC) linkage grant, LP110200321. In addition, Anagi Gamachchi is supported by an ``Australian Government Research Training Program Scholarship". 

\bibliographystyle{IEEEtranS}

\bibliography{ref_trustcom}

\end{document}